\documentclass[reprint,amsmath,amssymb,prb,superscriptaddress]{revtex4-2}
\usepackage{graphicx}
\usepackage{xcolor}
\usepackage{soul}
\usepackage{newtxtext} 
\usepackage[margin=0.9in]{geometry} 
\usepackage{graphicx}
\usepackage{dcolumn}
\usepackage{hyperref}
\usepackage{amsmath}
\usepackage{cleveref}
\usepackage{newtxmath}
\usepackage[utf8]{inputenc}

\hypersetup{colorlinks,
	linkcolor={blue!75!black!80!yellow},
	citecolor={blue!75!black!80!yellow},
	urlcolor={blue!75!black!80!yellow}
}

\makeatletter \renewcommand\@make@capt@title[2]{%
\@ifx@empty\float@link{\@firstofone}{\expandafter\href\expandafter{\float@link}}%
\sffamily{\textbf{#1}}\@caption@fignum@sep#2 }

\usepackage[normalem]{ulem}
\makeatletter
\renewcommand\@make@capt@title[2]{%
    \@ifx@empty\float@link{\@firstofone}{\expandafter\href\expandafter{\float@link}}%
    \sffamily{\textbf{#1}}\@caption@fignum@sep#2
}%

\begin{document}

\title{Generalized Design Principles for Hydrodynamic Electron Transport in Anisotropic Metals}

\author{Yaxian Wang}
\thanks{These authors contributed equally.}
\affiliation{John A. Paulson School of Engineering and Applied Sciences, Harvard University, Cambridge, MA 02138, USA}
\author{Georgios Varnavides}
\thanks{These authors contributed equally.}
\affiliation{John A. Paulson School of Engineering and Applied Sciences, Harvard University, Cambridge, MA 02138, USA}
\affiliation{Department of Materials Science and Engineering, Massachusetts Institute of Technology, Cambridge, MA 02139, USA}
\affiliation{Research Laboratory of Electronics, Massachusetts Institute of Technology, Cambridge, MA 02139, USA}
\author{Polina Anikeeva}
\affiliation{Department of Materials Science and Engineering, Massachusetts Institute of Technology, Cambridge, MA 02139, USA}
\affiliation{Research Laboratory of Electronics, Massachusetts Institute of Technology, Cambridge, MA 02139, USA}
\author{Johannes Gooth}
\affiliation{Max  Planck  Institute  for  Chemical  Physics  of  Solids,  N{\"o}thnitzer  Strasse  40,  01187  Dresden,  Germany}
\author{Claudia Felser}
\affiliation{Max  Planck  Institute  for  Chemical  Physics  of  Solids,  N{\"o}thnitzer  Strasse  40,  01187  Dresden,  Germany}
\author{Prineha Narang}
\email[Electronic address:\;]{prineha@seas.harvard.edu}
\affiliation{John A. Paulson School of Engineering and Applied Sciences, Harvard University, Cambridge, MA 02138, USA}

\begin{abstract}
\noindent Interactions of charge carriers with lattice vibrations, or phonons, play a critical role in unconventional electronic transport of metals and semimetals.
Recent observations of phonon-mediated collective electron flow in bulk semimetals, termed electron hydrodynamics, present new opportunities in the search for strong electron-electron interactions in high carrier density materials.
Here we present the general transport signatures of such a second-order scattering mechanism, along with analytical limits at the Eliashberg level of theory.
We study electronic transport, using \emph{ab initio} calculations, in finite-size channels of semimetallic ZrSiS and TaAs$_2$ with and without topological band crossings, respectively.  
The order of magnitude separation between momentum-relaxing and momentum-conserving scattering length-scales across a wide temperature range make both of them promising candidates for further experimental observation of electron hydrodynamics.
More generally, our calculations show that the hydrodynamic transport regime can be realized in a much broader class of anisotropic metals and does not, to first order, rely on the topological nature of the bands.
Finally, we discuss general design principles guiding future search for hydrodynamic candidates, based on the analytical formulation and our \emph{ab initio} predictions.
We find that systems with strong electron-phonon interactions, reduced electronic phase space, and suppressed phonon-phonon scattering at temperatures of interest are likely to feature hydrodynamic electron transport.
We predict that layered and/or anisotropic semimetals composed of half-filled $d$-shells and light group V/VI elements with lower crystal symmetry are promising candidates to observe hydrodynamic phenomena in future. 
\end{abstract}

\date{\today}
\maketitle

\section{Introduction}
Hydrodynamic electron transport, where charge carriers can flow collectively akin to a classical fluid, has recently garnered significant attention as a probe of strong electron interactions in conductors, with increasing technological relevance as electronic devices approach the micro- and nano-meter scale.
For instance, in a hydrodynamic conductor resistive processes occur predominantly at the boundaries, which alters the spatial distribution of Joule heating and can thereby significantly impact thermal design.
Further, it has been demonstrated that in a narrow conducting channel, collective flow can transfer charge more efficiently than the ballistic regime, thus achieving ``superballistic'' transport~\cite{bandurin2018fluidity}.
Microscopically, this requires that the total momentum of electrons is conserved, with the momentum-relaxing interactions of electrons with impurities, lattice vibrations, or the device boundary being significantly slower.
The hydrodynamic transport regime has until recently been inaccessible since at high temperatures the electron momentum is often relaxed by lattice vibrations, while at low temperatures by extrinsic scattering due to high impurity concentrations. 
In these systems charge transport is governed by diffusive processes, where the electrons lose their momentum after traveling an average ``mean free path'' distance ($l_{\rm mr}$). 

To observe hydrodynamic effects, the momentum conserving length scale ($l_{\rm mc}$) needs to dominate, necessitating electron-electron scattering to be frequent enough.
Since the strength of conventional Coulomb interactions depends inversely on the carrier density~\cite{Kotov2012electron}, hydrodynamic transport is most likely to occur in semiconductors or low carrier-density semimetals such as graphene. 
This is supported by the first observations of hydrodynamic electron flow in two dimensional electron gases (2DEG) (Al,Ga)As~\cite{jong1995hydrodynamic}, and more recently graphene~\cite{Bandurin2016negative,Crossno2016observation,KrishnaKumar2017superballistic}. 
However, potential applications and optimization of this transport regime are hindered because in semiconductors sufficiently high carrier concentrations are usually achieved by impurities, which in turn dominate the scattering at low temperatures; while in bulk metals with high carrier densities conventional Coulomb interactions are screened. 

The growth of high quality single crystals with very low levels of impurities has facilitated recent observations of hydrodynamic transport in a handful of high carrier-density systems such as semimetals WP$_2$~\cite{Gooth2018thermal} and WTe$_2$~\cite{vool2020imaging}, and delafossite metals PdCoO$_2$ and PtCoO$_2$~\cite{Nandi2018unconventional,moll2016evidence}, where the observation could not readily be explained using the Coulomb interaction~\cite{vool2020imaging,varnavides2021finite}.
Instead, a combination of theory and experiment shows an indirect electron-electron interaction, mediated by a virtual phonon, dominates at intermediate temperatures~\cite{vool2020imaging,varnavides2021finite}.
This provides an opportunity to realize and optimize hydrodynamic effects in a broader family of materials, ideally with high carrier mobility and tunable sample quality. 
Despite these theoretical advances in understanding the macroscopic observables of hydrodynamic phenomena~\cite{levitov_electron_2016,de_jong_hydrodynamic_1995,LucasKCFong2018hydrodynamic}, general design principles guiding the discovery of bulk hydrodynamic candidates remain elusive. 
In this \emph{Article}, we explore the phonon-mediated electron-electron scattering mechanism in anisotropic metals and semimetals with \emph{ab initio} calculations.
We investigate the possibility of hydrodynamic electron flow in semimetals ZrSiS and TaAs$_2$, the former of which has three-dimensional Dirac nodal lines while the later has no Dirac or Weyl crossings.
We find that both materials host strong phonon-mediated electron-electron interactions and are promising candidates to exhibit hydrodynamic behavior at relatively higher cryogenic temperatures than previously reported systems.
Inspired by the recent development of temperature-dependent imaging techniques that can spatially resolve the electron current profile via nitrogen-vacancy magnetometry~\cite{Ku2020,jenkins_imaging_2020,vool2020imaging}, we compute the current density profiles for various combinations of momentum relaxing and momentum conserving length scales, and characterize the resulting transport regimes at different temperatures to provide guidance for experimentally-relevant hydrodynamic observables.

Further, in this \emph{Article} we investigate the analytical limits of the phonon-mediated scattering lifetime, and provide empirical evidence to support these limits in light of this work and previously explored semimetals~\cite{coulter2018,Osterhoudt2021evidence,vool2020imaging}.
At temperatures significantly lower than the system's Debye temperature, when anharmonic phonon-phonon interactions are significantly slower than electron-phonon interactions, resulting in long-lived phonon-excitations which preferentially transfer their momenta to the electronic system, the second-order phonon-mediated electron lifetime is shorter than the first-order electron-phonon lifetime.
This effectively shifts the focus in the search for hydrodynamic candidates to ultra-pure materials with long momentum-relaxing mean free paths, significantly expanding the pool for future search.
Finally, we discuss the essential features leading to hydrodynamic electron transport in metallic candidates: 
(a) strong electron-phonon coupling, typically found in low-symmetry crystals composed of $d/p$ orbitals~\cite{butler1977electron}; 
(b) suppressed phonon-phonon interactions, e.g. realized through the ``acoustic bunching'' effect in systems with relatively large atomic mass difference~\cite{Lindsay2013first};
(c) high Fermi velocity often correlated with highly dispersive electronic bands as well as low levels of disorder.
From these predictions and observations we conclude that anisotropic quantum materials composed of transition metals and group V/VI elements host multiple materials families suitable for exhibiting and optimizing hydrodynamic electron flow.

\section{\label{sec:overview_materials}Overview of material properties in phonon-mediated hydrodynamic candidates}

At first glance, the observation of hydrodynamic electron transport in three dimensional bulk conductors appears serendipitous in nature, since conventional electron-electron Coulomb interactions are expected to be short-ranged due to electron screening effects.
More specifically, recent temperature- and spatially-resolved measurements on WTe$_2$ show evidence of non-uniform flow at $\sim20$~K~\cite{vool2020imaging}. 
Meanwhile, bulk transport signatures on WP$_2$ feature strong deviations from the Wiedemann-Franz law~\cite{Gooth2018thermal}, while the temperature-dependence of phonon linewidths in WP$_2$ cannot be readily explained via lattice anharmonicity models and provides evidence of strong phonon-electron coupling~\cite{Osterhoudt2021evidence}.
Similarly, the width-dependent resistivity in narrow conducting channels of PdCoO$_2$ indicated a $l_{\rm mc}/l_{\rm mr}\sim 0.1$ ratio meaning momentum-conserving mean free paths are as short as a few microns~\cite{moll2016evidence}, in contrast to conventional electron-electron interactions from the Fermi-liquid theory estimating $l_{\rm mc}$ on the order of thousands of microns~\cite{Giuliani1982lifetime}.
In a series of recent theoretical predictions, phonon-mediated electron-electron scattering is shown to be important for the non-diffusive electron transport in these systems~\cite{coulter2018,Gooth2018thermal,Osterhoudt2021evidence,vool2020imaging,varnavides2021finite}.
Since momentum-relaxing scattering needs to be minimized to observe hydrodynamic effects, it is important for material candidates to exhibit long momentum-relaxing $l_{\mathrm{mr}}$.
In this section we examine the metals and semimetals WP$_2$, WTe$_2$, and PdCoO$_2$ that exhibit signatures of hydrodynamics and the material properties that enable this physics.

\begin{figure*}
    \centering
    \includegraphics[width=0.95\linewidth]{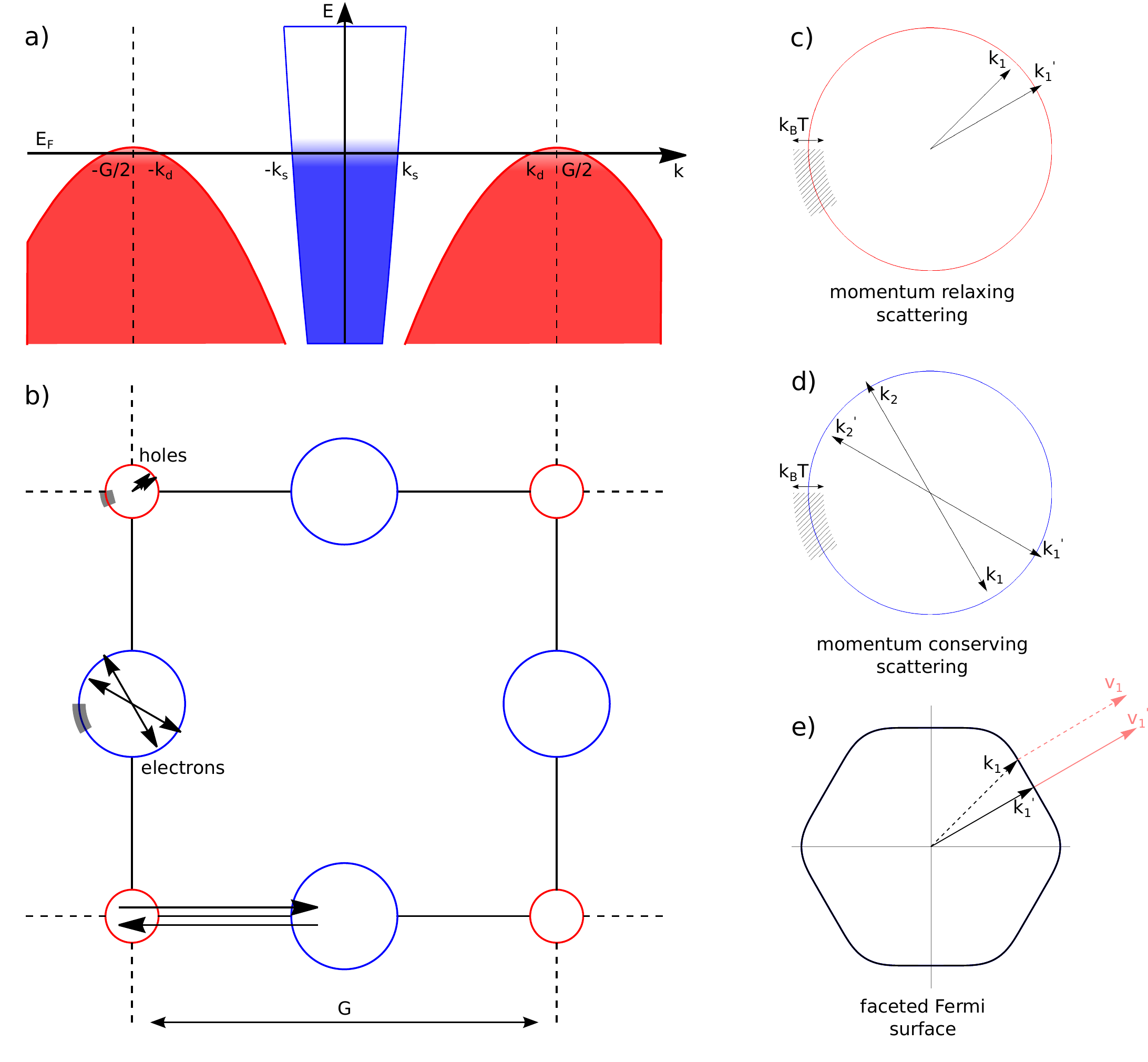}
    \caption{(a) Schematic energy dispersion diagram along a high symmetry direction for a semimetal, with highly-dispersive $s/p$ electron-like (blue) and ``heavy'' $d/f$ hole-like (red) bands crossing the Fermi energy.
    Due to the separation of the Fermi pockets, lattice excitations with a large wave numbers are needed for transition between $k_s$ and $k_d$.
    (b) Fermi surface diagram of the semimetal shown in (a) across the first Brillouin zone, showing intra-pocket (top-left) and inter-pocket (bottom) scattering events.
    The square lattice Brillouin zone, with reciprocal lattice vector $G$, is shown in solid lines.
    Adapted from Ref.~\citenum{kaveh1984electron}.
    (c) Electron-phonon momentum relaxing scattering event, where initial electronic state $\boldsymbol{k}$ is scattered into final state $\boldsymbol{k'}$, exchanging both energy and momentum to the phonon mode $\boldsymbol{q}=\boldsymbol{k'}-\boldsymbol{k}$.
    The thermally accessible states with energy $\sim k_BT$ around the Fermi energy are indicated by the shaded area.
    (d) Electron-electron momentum conserving scattering event, where two initial electronic states $\boldsymbol{k_1}$ and $\boldsymbol{k_2}$ are scattered into final states $\boldsymbol{k_1'}$ and $\boldsymbol{k_2'}$.
    The scattering process conserves both energy and momentum.
    (e) A faceted Fermi surface, on which a scattering event between wave vectors from $\boldsymbol{k}$ to $\boldsymbol{k'}$ may lead to negligible change in the electron velocity such that the total momentum stays ``quasi'' conserved.}
    \label{fig:schematics}
\end{figure*}

\begin{figure*}
    \centering
    \includegraphics[width=1\linewidth]{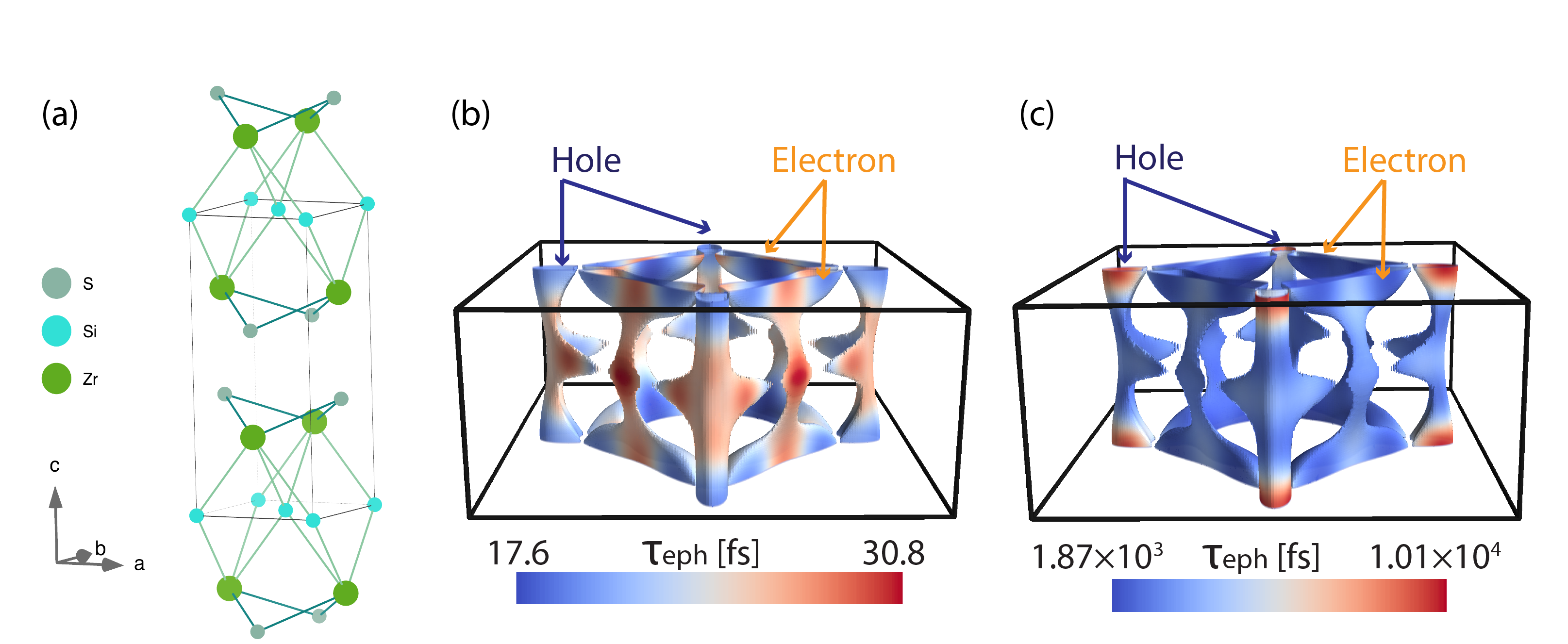}
    \caption{(a) Crystal lattice of ZrSiS, highlighting the layered structure. 
    (b-c) Electron-phonon lifetimes ($\tau_{\rm eph}$) at (b)~298~K and (c)~4~K projected on the Fermi surface of ZrSiS. 
    Similarly, while both electron and hole pockets have long lived carriers, lowering temperature quenches the scattering events on the hole pockets much more significantly.}
    \label{fig:ZrSiS_FS}
\end{figure*}

Intuitively, delocalized electrons, for example from $s$ orbitals, usually lead to large dispersion in band energy, that is, larger Fermi velocity, $v_F$.
However, they typically- result in symmetric bonding states often found in isotropic lattices making them less sensitive to lattice vibrations.
To access the second-order phonon-mediated interactions, we seek strong electron-phonon coupling across the Fermi surface (see ~\Cref{sec:eph-coupling}), therefore more localized electrons, for example hybridized $d$ orbitals, are advantageous. 
By contrast, suppression of the first-order momentum-relaxing scattering can be achieved in a number of ways~\cite{kaveh1984electron,wilson1938electrical} (each individually effective though a confluence would be beneficial):
I) In semimetals with well separated electron/hole Fermi pockets across the Brillouin zone. 
Crystals with lower symmetry and complex orbital hybridization usually have their conduction band minimum and valance band maximum shifted from the zone center, so their Fermi surfaces are located away from each other.
Here, at low temperatures, phonon excitations lack the momentum to couple electrons from one pocket to another, and only intra-band scattering is allowed  (\cref{fig:schematics}a-b);
II) At low temperatures where the phonon phase-space is significantly reduced,  momentum-relaxing scattering is ineffective at relaxing the system's momentum appreciably (\cref{fig:schematics}c), while momentum-conserving electron-electron scattering via the `instantaneous' emission and re-absorption of phonon modes (\cref{fig:schematics}d) is allowed to take place for larger $\boldsymbol{q}$;
III) On Fermi surfaces with non-trivial geometry where small scattering efficiency factor leads to ``quasi'' conservation of electron momentum (\cref{fig:schematics}e).
Specifically, at very low temperatures, small phonon momenta $\boldsymbol{q}$ always result in a small angle between the incident, $\boldsymbol{k}$,  and scattered electron momentum, $\boldsymbol{k'}=\boldsymbol{k}-\boldsymbol{q}$.
On a Fermi surface with non-trivial geometry, the electron velocity $v_F$ can deviate significantly from the wave vector and thus affect the transport-relevant scattering efficiency~\cite{varnavides2021finite,van2020sondheimer}.
We can quantify this effect by taking into account the scattering angle after scattering~\cite{coulter2018,van2020sondheimer}:
\begin{equation}
    1-\mathrm{cos}\theta = 1-\frac{v_{n\boldsymbol{k}}\cdot v_{m\boldsymbol{k+q}}}{|v_{n\boldsymbol{k}}||v_{m\boldsymbol{k+q}}|}.
    \label{eq:scatteringefficiency}
\end{equation}
One such example is shown in \cref{fig:schematics}e, when the concave portions of the Fermi surface, result in a strong velocity and mean free path distribution anisotropy~\cite{ong1991geometric,wang2020anisotropic}, and can further reduce the momentum relaxing scattering efficiency.

Recent first principles calculations have provided significant insights into the state-resolved momentum relaxing lifetimes in these systems~\cite{coulter2018,vool2020imaging,varnavides2021finite}. 
WP$_2$~\cite{coulter2018} and WTe$_2$~\cite{vool2020imaging} both show a similar trend in the carrier lifetime distribution on their Fermi surfaces. 
At high temperatures, the momentum relaxing lifetimes are distributed more evenly on electron/hole pockets compared to extremely low temperatures, where the long lived electrons are ``focused'' to specific spots on the hole pockets, which have open Fermi surface shape and thus reduced electronic phase space. 
For PdCoO$_2$, which has a faceted open Fermi surface in a hexagonal shape, at high temperatures, long-lived electrons span the entire Brillouin zone along both $k_z$ and $k_{x/y}$~\cite{varnavides2021finite}; while at 4~K, the long-lived electrons are located only at the corners of the hexagonal Fermi surface at $k_z=0$.
Taken together, the highly dispersive hole bands give rise to large Fermi velocity and smaller phase space for electron-phonon scattering. 
Yet, the strong electron-phonon coupling arising from $d$-orbitals, leads to a strong second-order electron-electron interaction mediated by a virtual phonon.
These emerging common features  motivate us to search for hydrodynamic candidates in layered and/or anisotropic structures with transition metals and group V/VI elements, and to postulate generalized design principles.
Meanwhile, various electronic transport signatures are indicative of the, otherwise hard to experimentally obtain, electron-phonon coupling strength. 
For example, large RRR (residual-resistivity ratio) is an important sign of low-impurity samples, and large MR (magnetoresistance) can indicate charge compensation, unconventional Fermi surface topology and scattering mechanisms.

\section{\label{sec:abinitio_prediction}Ab-initio Predictions of Electron Hydrodynamics in anisotropic metals}
Following the design principles discussed in Sec.~\ref{sec:overview_materials}, we investigate layered compounds with transition metals that show unconventional transport signatures, more specifically MR.
This is because while hydrodynamic electron flow has only been experimentally reported in a few bulk systems, anomalous MR has been heavily studied in recent literature. 
Large positive (and mostly linear) MR has been found in many semimetals including but not limited to WTe$_2$~\cite{Ali2014WTe2,pletikosi2014electronic}, LaBi~\cite{kumar2016observation}, NbIrTe$_4$~\cite{Zhou2019nonsaturating} etc.
In addition, large extremely anisotropic MR is reported in Dirac nodal line semimetal ZrSiS~\cite{Ali2016butterfly,Lv2016extremely} and ZrSi(Se,Te)~\cite{chiu2019origin}.
Further, negative MR was found in topological semimetals~\cite{Zhang2016Bell,Arnold2016negative,Li2016negative}, a semimetal TaAs$_2$ without Dirac dispersion~\cite{Luo2016}, and at the LaAlO$_3$/SrTiO$_3$ interface~\cite{diez2015giant}. 
These observations suggest the pool of candidates for hydrodynamic electron flow in bulk materials may be much larger than previously thought.
In this section we present calculations of ZrSiS and TaAs$_2$ as case studies to evaluate these design principles. 

\begin{figure*}
    \centering
    \includegraphics[width=1\linewidth]{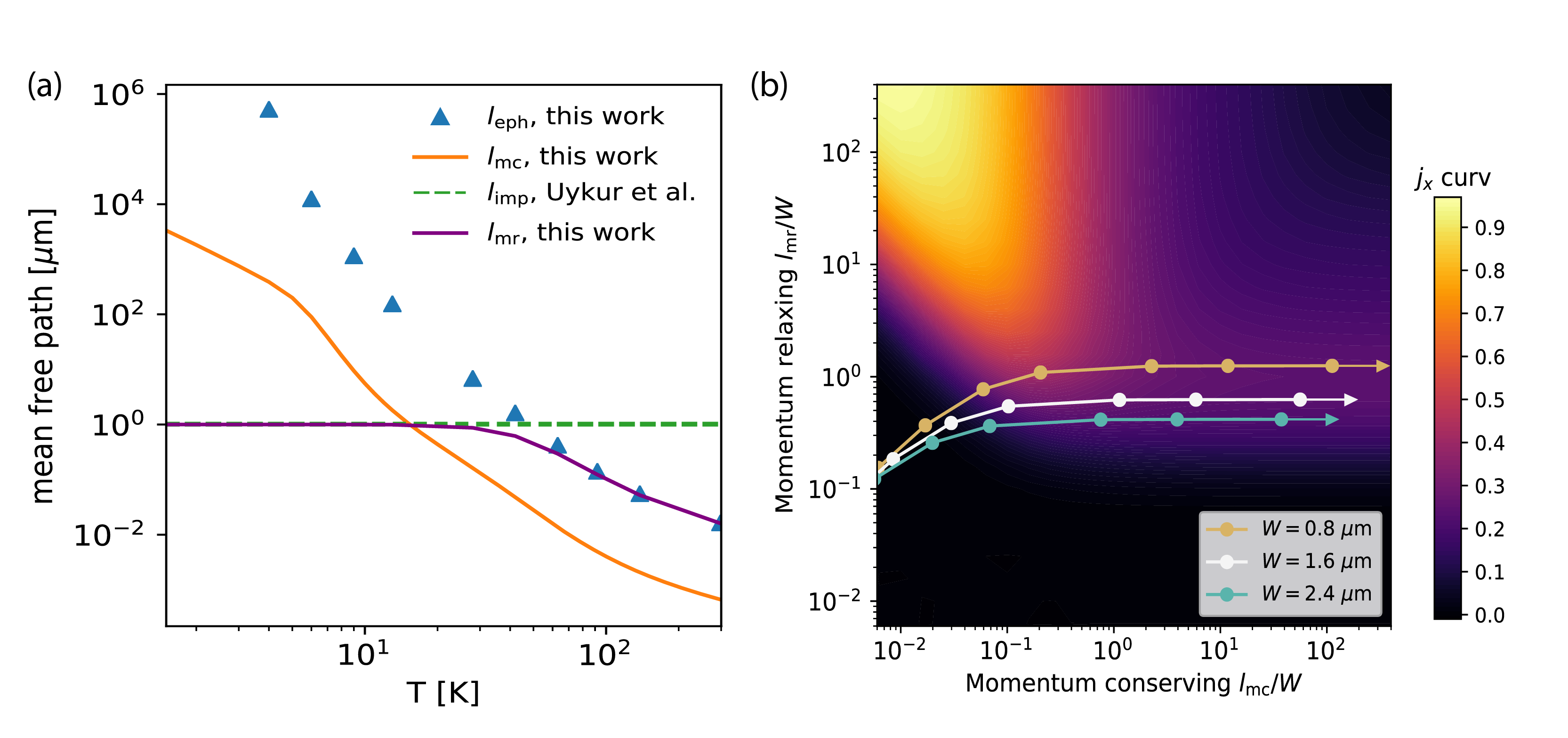}
    \caption{(a) Temperature dependence of momentum relaxing electron-phonon ($l_{\rm mr}$) and momentum conserving electron-electron ($l_{\rm mc}$) mean free paths in ZrSiS calculated from \emph{ab initio}.
    The impurity mean free path is taken as 1~$\mu{\rm m}$ from Uykur \emph{et al.}~\cite{Uykur2019magneto}.
    (b) Normalized $j_x$ curvature phase diagram.
    Overlaid lines show the trajectories with the decreasing temperature with different channel widths.}
    \label{fig:ZrSiS_phasediagram}
\end{figure*}

ZrSiS crystallizes in the PbFCl-type tetragonal $P4/nmm$ structure (\cref{fig:ZrSiS_FS}a), and has been shown to host a three-dimensional Dirac line node and feature a chiral anomaly and extremely large non-saturating MR~\cite{Schoop2016dirac,Fu2019,Hu2017nearly,Ali2016butterfly,Singha2017large,Lv2016extremely}. 
Non-saturating MR results from charge compensation of its high mobility carriers with a ratio of $\sim0.94$ electrons to holes, similar to WTe$_2$. 
Moreover, the tube-shaped Fermi surface oriented along the crystallographic $\hat{c}$ direction leads to open electron orbits under magnetic fields, which can give rise to strong anisotropic MR~\cite{Lv2016extremely,Singha2017large}. 
These signatures indicate the possibility hydrodynamic electron flow in this material that we investigate next.

\Cref{fig:ZrSiS_FS}b-c shows the Fermi surface of ZrSiS, composed of four degenerate electron and hole pockets due to the mirror symmetry in the $ab$ plane all in open shape, in agreement with previous work~\cite{Schoop2016dirac,Lv2016extremely,Fu2019,Ali2016butterfly}.
The behavior of momentum-relaxing electron-phonon lifetimes ($\tau_{\rm eph}$) at 298~K and 4~K shows substantial similarity with those in other systems mentioned above, such as WTe$_2$ and WP$_2$. 
While at 298~K longer-lived carriers appear evenly distributed between electron and hole pockets, at 4~K the only long-lived carriers are located on hole pockets at the zone boundary. 
Using the formalism developed in Sec.~\ref{sec:eph-coupling}, the phonon-mediated electron-electron lifetime ($\tau_{\rm ee(ph)}$) is much shorter than the momentum relaxing electron-phonon lifetimes across all temperatures (\cref{fig:ZrSiS_phasediagram}a).
Therefore, the crossover between the dominant scattering mechanism is determined by the impurity scattering mean free path ($l_{\mathrm{imp}}$).
High quality ZrSiS samples have been shown to have an impurity mean free path $l_{\mathrm{imp}}>1\mu{\rm m}$~\cite{Uykur2019magneto,Schilling2017flat}.
Using $l_{\mathrm{imp}}=1\mu{\rm m}$, momentum relaxing events appear to dominate below $\sim20$~K (\cref{fig:ZrSiS_phasediagram}a), allowing for the possibility of observing hydrodynamic flow at higher temperatures.
Further, the expected current density ($j_x$) profiles with various combinations of $l_{\rm mc}$ and $l_{\rm mr}$ in a narrow conducting channel are computed from solving the spatially-resolved Boltzmann transport equation (BTE), discussed in our prior work~\cite{vool2020imaging,varnavides2021finite}.
The contour plot of the $j_x$ curvature gives a general metric to characterize different transport regime limits. 
Using the $l_{\rm mc}$ and $l_{\rm mr}$ values obtained from \emph{ab initio} calculations, we can bridge the microscopic scattering mechanisms with these observables and examine how strong the hydrodynamic effect is in samples with different dimensions (\cref{fig:ZrSiS_phasediagram}b).
Our results suggest hydrodynamic electron flow can be realized in narrow, sub-micron devices.
Alternatively, if sample quality is further improved to $l_{\mathrm{imp}}\approx5\mu{\rm m}$, the hydrodynamic window can be expanded to lower temperatures and wider devices. 
Compared to that of WTe$_2$, $l_{\mathrm{mc}}$ is relatively small in ZrSiS, due to a larger carrier concentration ($\sim9\times10^{19}{\rm cm}^{-3}$ in ZrSiS~\cite{Ali2016butterfly} versus $\sim2\times10^{19}{\rm cm}^{-3}$ in WTe$_2$~\cite{vool2020imaging}).

\begin{figure*}
    \centering
    \includegraphics[width=1\linewidth]{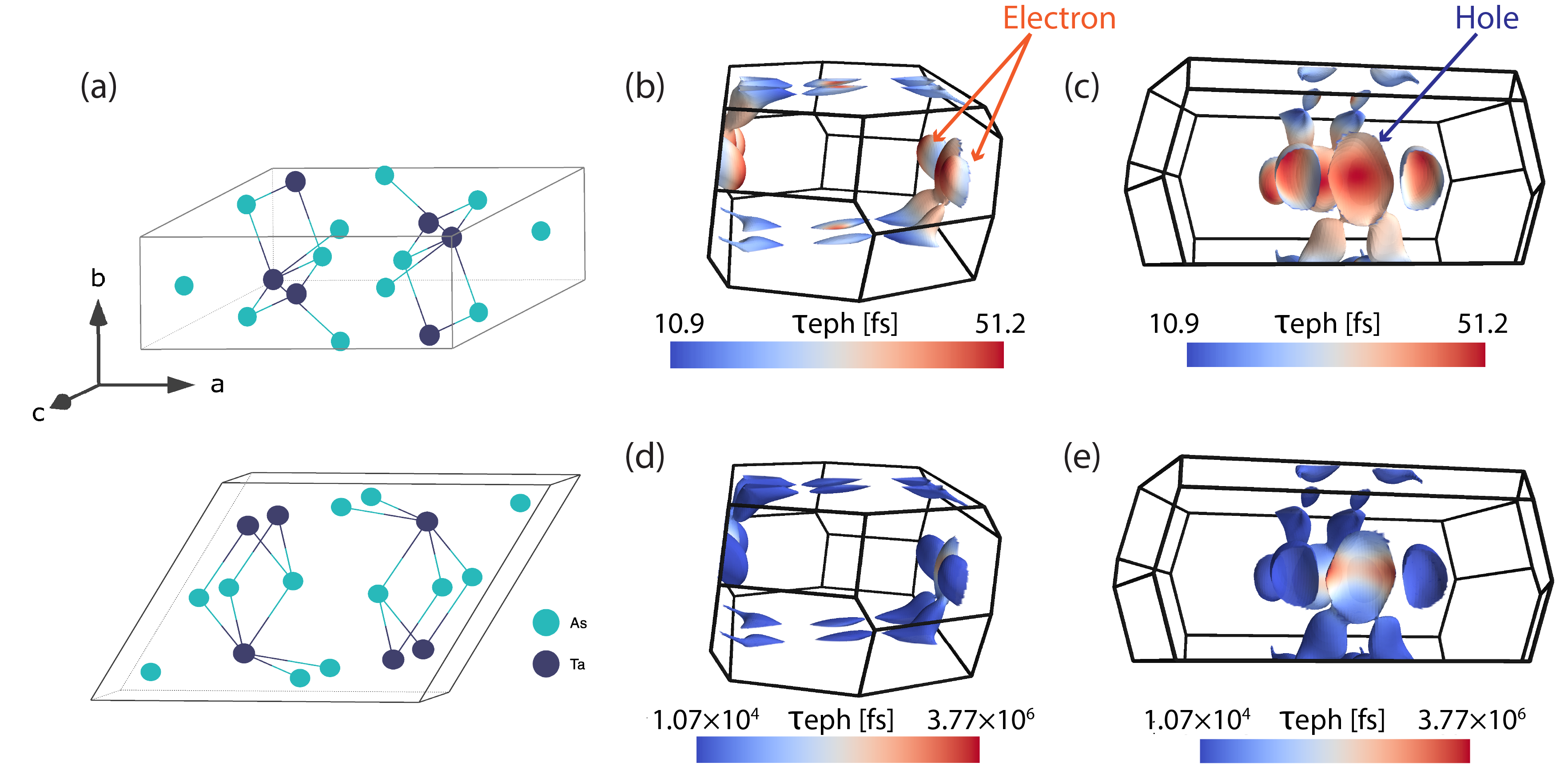}
    \caption{(a) Crystal lattice of TaAs$_2$, highlighting the monoclinic structure. 
    (b-c)~Electron-phonon lifetimes ($\tau_{\rm eph}$) at 298~K and (d-e)~4~K projected on the Fermi surface of TaAs$_2$. 
    Different views are presented due to the low symmetry Brillouin zone and the complexity of the electron and hole pockets.
    The features are highlighted that hole pockets at low temperatures feature much longer momentum relaxing lifetimes.}
    \label{fig:TaAs2_FS}
\end{figure*}

While hydrodynamic signatures in the topological semimetals we considered so far (ZrSiS,  WP$_2$, and WTe$_2$) are similar, it is intriguing to address the necessity of topologically protected states in the context of hydrodynamic transport. 
The role of topology in Dirac-Weyl semimetals that exhibit signatures of hydrodynamics has sparked intense debates in the field.
To address this open question, we turn to TaAs$_2$ as a case study for a semimetal without nontrivial band crossings. 
TaAs$_2$ crystallizes in the monoclinic $C12/m1$  structure with two chemical sites for As atoms, one forming Ta-As planes with Ta atoms and the other bridging the interlayer coupling (\cref{fig:TaAs2_FS}a).
As a non magnetic material, TaAs$_2$ has been shown to have giant MR from high mobility compensated charge carriers~\cite{butcher2019fermi,Luo2016,Wu2016giant,wang2016resistivity,yuan2016large}.
However, the band crossings are gapped in the presence of spin orbit coupling, making it an interesting case to study hydrodynamic behavior in topologically trivial semimetals.
Owing to a small effective mass $m^*\sim 0.3m_0$~\cite{butcher2019fermi,wang2016resistivity}, the carrier mean free path is estimated to be $\sim 10\mu{\rm m}$ at 2~K, significantly longer than that of WTe$_2$ and ZrSiS.
Moreover, in a high quality TaAs$_2$ crystal with carrier density $\approx 2.8\times 10^{18}$~cm$^{-3}$, the MR is nearly one magnitude larger than that of WTe$_2$~\cite{Wu2016giant}, and the carrier mobility is up to $\approx 1.2\times 10^{5}$~cm$^2$/V/s~\cite{Luo2016}.
More importantly, the quadratic field dependence of its MR obeys the semiclassical model, representative of conventional (semi)metals.

These observations motivate us to examine in further detail the temperature dependent mean free paths for both momentum relaxing and momentum conserving events, shown in~\cref{fig:TaAs2_phasediagram}a. 
Similar to ZrSiS, the phonon mediated electron-electron mean free path is shorter than the first-order electron-phonon mean free path for all temperature ranges studied.
Combined with a long impurity mean free path, this provides a wide temperature window, as well as a wide range of channel widths in which to expect hydrodynamic flow (\cref{fig:TaAs2_phasediagram}b).
Compared to ZrSiS (Fig.~\ref{fig:ZrSiS_phasediagram}b) and WTe$_2$~\cite{vool2020imaging}, hydrodynamic flow in TaAs$_2$ appears more pronounced, calling for experimental verification.

Investigating the momentum relaxing lifetime distribution on the Fermi surface of TaAs$_2$ (\cref{fig:TaAs2_FS}b-e), we find similar features to those discussed before.
TaAs$_2$ has the valence band maximum located at the L point while the conduction band minimum is shifted between $\Gamma$ and L, leading to the coexistence of a few pairs of electron/hole Fermi surfaces located at the boundary of the Brillouin zone.
The electron band forms a closed Fermi surface while the hole band forms an open Fermi surface~\cite{butcher2019fermi}.
At high temperatures, the distribution of $\tau_{\rm eph}$ lifetimes is narrow and fairly evenly distributed between the electron and hole pockets.
At low temperatures, however, the electron lifetimes are considerably shorter than those of hole carriers, consistent with other semimetals discussed in this work.
Despite attempts to classify TaAs$_2$ as a new kind of topological material with $\mathbb{Z}_2$ invariant (0;111) from density functional theory calculations~\cite{Luo2016}, the absence of linear Dirac or Weyl band dispersion indicates that topologically protected band crossings are not a salient ingredient for observing hydrodynamic electron flow.
The possibility of observing hydrodynamic flow in TaAs$_2$ is intriguing, suggesting that strong phonon-mediated electron-electron interactions can be found in topologically trivial metals.
\section{Electron-Phonon Coupling Interactions and the analytical limits \label{sec:eph-coupling}}
Our \textit{ab initio} calculations suggest that, in materials with strong electron-phonon interactions, $l_{\mathrm{mc}} < l_{\mathrm{mr}}$ across all temperatures of interest.
In this section, we seek to understand this observation by comparing the second-order phonon-mediated electron-electron interaction analytically to the first-order electron-phonon interaction.
To first order in the atomic displacements, the coupled electron-phonon system is described by the Hamiltonian~\cite{giustino2016}:
\begin{align}
    &\hat{H} = \sum_{n\boldsymbol{k}}\epsilon_{n\boldsymbol{k}}\hat{c}^{\dagger}_{n\boldsymbol{k}}\hat{c}_{n\boldsymbol{k}} + \sum_{\boldsymbol{q}\nu}\hbar \omega_{\boldsymbol{q}\nu}\left(\hat{a}^{\dagger}_{\boldsymbol{q}\nu}\hat{a}_{\boldsymbol{q}\nu} + 1/2 \right) \notag \\
    &\; + N^{-1/2}\sum_{\boldsymbol{k}\boldsymbol{q}mn\nu}g_{mn\nu}(\boldsymbol{k},\boldsymbol{q})\hat{c}_{m\boldsymbol{k+q}}^{\dagger}\hat{c}_{n\boldsymbol{k}}\left(\hat{a}_{\boldsymbol{q}\nu}+\hat{a}^{\dagger}_{-\boldsymbol{q}\nu}\right),
\end{align}
where $\epsilon_{n\boldsymbol{k}}$ is the single-particle energy for an electron with crystal momentum $\boldsymbol{k}$ in band $n$, $\omega_{\boldsymbol{q}\nu}$ is the frequency for a phonon with crystal momentum $\boldsymbol{q}$ and polarization $\nu$, $\hat{c}^{\dagger}_{n\boldsymbol{k}}$ and $\hat{c}_{n\boldsymbol{k}}$ ($\hat{a}^{\dagger}_{\boldsymbol{q}\nu}$ and $\hat{a}_{\boldsymbol{q}\nu}$) are the fermionic (bosonic) creation and annihilation operators respectively, $g_{mn\nu}(\boldsymbol{k},\boldsymbol{q})$ is the matrix element coupling electrons and phonons, and $N$ is the number of unit cells in the phonon supercell.

Using Fermi's golden rule to first and second order respectively, we can arrive at expressions for the inverse lifetimes of the electronic system due to the coupling with phonons~\cite{coulter2018}:
\begin{align}
    \frac{1}{\tau_{n\boldsymbol{k}}^{\mathrm{eph}}} &= 
    \frac{2\pi}{\hbar}\sum_{m\nu\pm}\int \frac{d\boldsymbol{q}}{\Omega_{\mathrm{BZ}}}
    \delta\left(\epsilon_{m\boldsymbol{k+q}}-\epsilon_{n\boldsymbol{k}} \mp \hbar \omega_{\boldsymbol{q}\nu}\right) \notag \\ 
    &\times \left[n_{\boldsymbol{q}\nu}+\frac{1}{2} \mp \left(\frac{1}{2}-f_{m\boldsymbol{k+q}}\right)\right] \left| g_{mn\nu}(\boldsymbol{k},\boldsymbol{q})\right|^2 \label{eq:tau-eph} \\
    \frac{1}{\tau_{n\boldsymbol{k}}^{\mathrm{ee(ph)}}} &= 
    \frac{2\pi}{\hbar}\sum_{mpr} \int \frac{d\boldsymbol{q}}{\Omega_{\mathrm{BZ}}} \int \frac{d\boldsymbol{k'}}{\Omega_{\mathrm{BZ}}} \left| M_{nmpr}(\boldsymbol{k},\boldsymbol{k'},\boldsymbol{q})\right|^2 \notag \\
    &\times \left[f_{p\boldsymbol{k'}}f_{m\boldsymbol{k+q}} + f_{r\boldsymbol{k'+q}}\left(1-f_{p\boldsymbol{k'}}-f_{m\boldsymbol{k+q}}\right)\right] \notag \\
    &\times \delta\left(\epsilon_{n\boldsymbol{k}}+\epsilon_{r\boldsymbol{k'+q}} -\epsilon_{p\boldsymbol{k'}}-\epsilon_{m\boldsymbol{k+q}}\right) \label{eq:tau-eeph},
\end{align}
where $f$ and $n$ are the Fermi-Dirac and Bose-Einstein equilibrium distribution functions of electrons and phonons respectively, and $\Omega_{\mathrm{BZ}}$ is the volume of the first Brillouin zone.
\Cref{eq:tau-eph,eq:tau-eeph} describe the momentum-relaxing electron-phonon $\left(\tau_{n\boldsymbol{k}}^{\mathrm{e-ph}}\right)$ and momentum-conserving phonon-mediated electron-electron $\left(\tau_{n\boldsymbol{k}}^{\mathrm{ee(ph)}}\right)$ interactions at first and second order respectively, with
\begin{align}
    M_{nmpr}(\boldsymbol{k},\boldsymbol{k'},\boldsymbol{q}) = \sum_{\nu} \frac{g_{mn\nu}(\boldsymbol{k},\boldsymbol{q})^* g_{pr\nu}(\boldsymbol{k'},\boldsymbol{q})}{\hbar \omega_{\boldsymbol{q\nu}}+\epsilon_{n\boldsymbol{k}}-\epsilon_{m\boldsymbol{k+q}}+i \eta}.
\end{align}
Since we typically only consider interactions near the Fermi level, we introduce Fermi-surface averaged versions of~\cref{eq:tau-eph,eq:tau-eeph} at the Eliashberg level of theory~\cite{Allen1978,coulter2018}:
\begin{align}
    \tau_{\rm eph}^{-1} &= \frac{\pi \beta}{2 g(\epsilon_F)} \sum_{\nu} \int \frac{d\boldsymbol{q}}{\Omega_{\mathrm{BZ}}} G_{\boldsymbol{q}\nu} \frac{\omega_{\boldsymbol{q}\nu}}{\sinh^2(\hbar \beta \omega_{\boldsymbol{q}\nu}/2)} \label{eq:tau-eph-2} \\
    \tau_{\rm ee(ph)}^{-1} &= \frac{2 \pi}{\hbar g(\epsilon_F)} \sum_{\nu} \int \frac{d\boldsymbol{q}}{\Omega_{\mathrm{BZ}}} G_{\boldsymbol{q}\nu}^2 \gamma(\hbar \beta \Bar{\omega}_{\boldsymbol{q} \nu}) \label{eq:tau-eeph-2}.
\end{align}
Here, $\beta = 1/k_B T$ is the dimensionless inverse temperature, $g(\epsilon_F)$ is the density of states at the Fermi level per unit cell, and $G_{\boldsymbol{q}\nu}$ is the dimensionless Fermi-surface integrated electron-phonon coupling strength:
\begin{align}
    G_{\boldsymbol{q}\nu} = \sum_{nm} \int \frac{g_s d\boldsymbol{k}}{\Omega_{\mathrm{BZ}}} \left| g_{mn\nu}(\boldsymbol{k},\boldsymbol{q})\right|^2 \delta(\epsilon_{n\boldsymbol{k}}-\epsilon_F) \delta(\epsilon_{m\boldsymbol{k+q}}-\epsilon_F), 
    \label{eq:g-coupling}
\end{align}
and $\gamma(x)$ is the complex-valued integral, evaluated at the complex phonon frequency:
\begin{align}
    \gamma(x) &\equiv \int_{-\infty}^{\infty} dy \frac{1}{4} \frac{y^2}{\sinh^2(y/2)|x-y|^2} \label{eq:gamma}\\
    \Bar{\omega}_{\boldsymbol{q} \nu} &= \omega_{\boldsymbol{q} \nu} + i \tau_{\boldsymbol{q} \nu}^{-1}/2 = \omega_{\boldsymbol{q} \nu} + i \pi \omega_{\boldsymbol{q} \nu} G_{\boldsymbol{q} \nu} \notag .
\end{align}
\Cref{eq:tau-eph-2,eq:tau-eeph-2} can now directly be compared, by seeking a suitable function approximation to $\gamma(x)$.
To this end, we express the integrand of~\cref{eq:gamma} as a power series near $\hbar \beta  \omega_{\boldsymbol{q}\nu}$ using a (0,2) Pad\'e approximant:
\begin{widetext}
\begin{align}
    \gamma(\hbar \beta \bar{\omega}_{\boldsymbol{q}\nu}) &\approx \int_{-\infty}^{\infty} dy \left(\hbar \beta \omega_{\boldsymbol{q}\nu} \right)^2 / \notag \\
    &\left\{
    \cosh (\hbar \beta \omega_{\boldsymbol{q}\nu} ) \left[2y^2 +6\left(\pi G_{\boldsymbol{q} \nu} y\right)^2 + \left(\pi G_{\boldsymbol{q} \nu} y\hbar \beta \omega_{\boldsymbol{q}\nu}\right)^2 
    - 2 y\hbar\beta\omega_{\boldsymbol{q}\nu} \left(2+8\left(G_{\boldsymbol{q} \nu} \pi\right)^2+\left(G_{\boldsymbol{q} \nu}\pi \hbar \beta \omega_{\boldsymbol{q}\nu}\right)^2\right)\right.\right. \notag \\
    &\qquad \left.\left. 
    + \left(\hbar\beta\omega_{\boldsymbol{q}\nu}\right)^2 \left(2+12\left(G_{\boldsymbol{q} \nu} \pi\right)^2+\left(G_{\boldsymbol{q} \nu}\pi \hbar \beta \omega_{\boldsymbol{q}\nu}\right)^2\right) \right] 
    -2 \left(6 \pi ^2 G_{\boldsymbol{q} \nu}^2+1\right) \hbar \beta \omega_{\boldsymbol{q} \nu} ^2
    -2 \left(3 \pi ^2  G_{\boldsymbol{q} \nu}^2+1\right) y^2\right. \notag \\
    &\qquad\left. + 4 \left(1+4 \pi ^2 G_{\boldsymbol{q} \nu}^2\right) y \hbar \beta \omega_{\boldsymbol{q}\nu} 
    -2 \pi ^2 G_{\boldsymbol{q} \nu}^2 \hbar \beta \omega_{\boldsymbol{q} \nu} (\hbar \beta \omega_{\boldsymbol{q}\nu} -y) (3 \hbar \beta \omega_{\boldsymbol{q}\nu} -2 y) \sinh (\hbar \beta \omega_{\boldsymbol{q}\nu} )\right\} \label{eq:pade-approximant}
\end{align}

\begin{figure*}
    \centering
    \includegraphics[width=\linewidth]{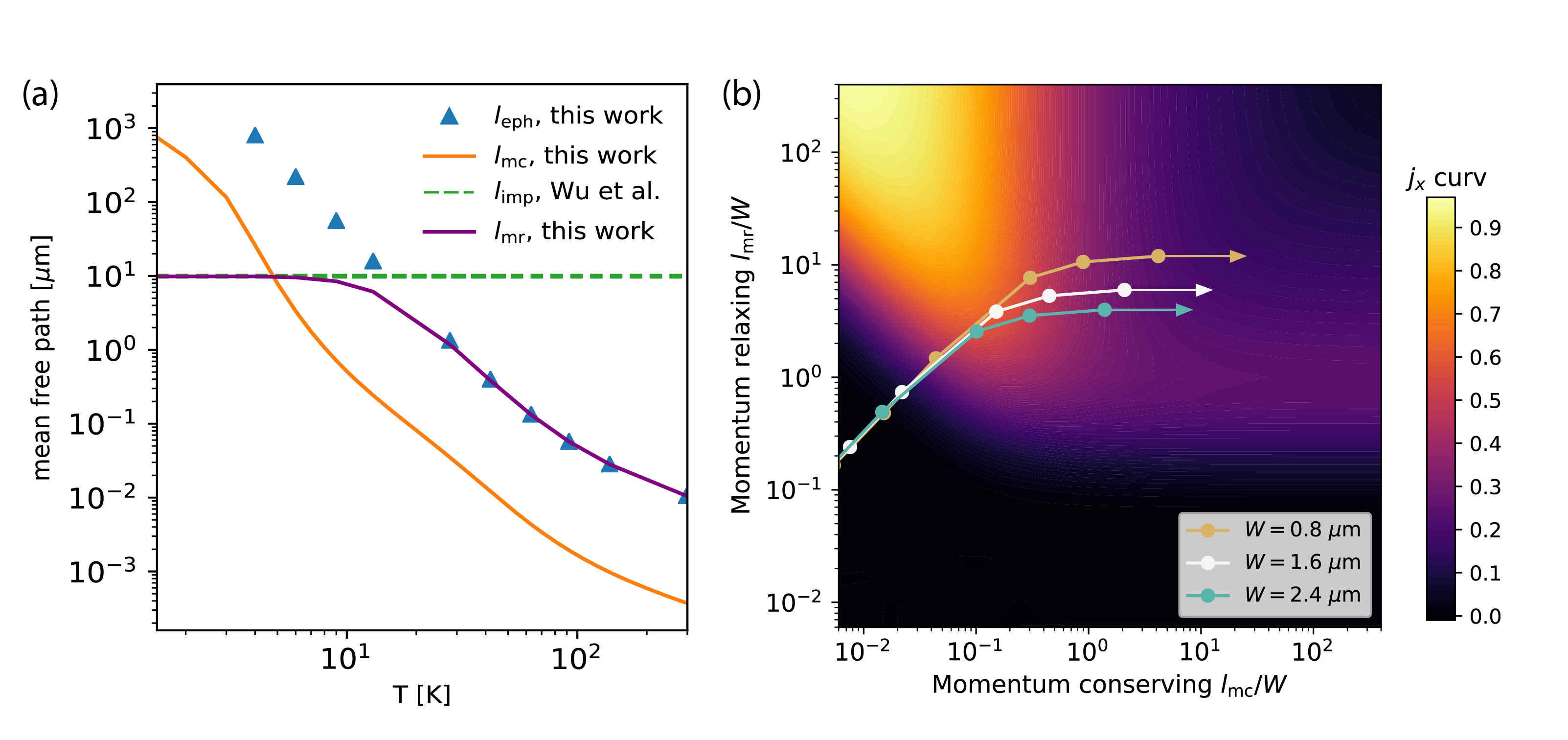}
    \caption{
       (a) Temperature dependence of momentum relaxing electron-phonon ($l_{\rm mr}$) and momentum conserving electron-electron ($l_{\rm mc}$) mean free paths in TaAs$_2$ calculated from \emph{ab initio}.
    The impurity mean free path is estimated as 10~$\mu{\rm m}$ from earlier works~\cite{butcher2019fermi,Luo2016}.
    (b) Normalized $j_x$ curvature phase diagram.
Overlaid lines show the trajectories with the decreasing temperature with different channel width. }
    \label{fig:TaAs2_phasediagram}
\end{figure*}

Applying the fundamental theorem of calculus to the antiderivative of~\cref{eq:pade-approximant} and simplifying the expression we obtain:
\begin{align}
    \gamma(\hbar \beta \bar{\omega}_{\boldsymbol{q}\nu}) &\approx 
    \frac{\hbar \beta \omega_{\boldsymbol{q}\nu}}{2 G_{\boldsymbol{q} \nu} \sinh^{\frac{3}{2}}\left(\frac{\hbar \beta \omega_{\boldsymbol{q}\nu}}{2}\right) \sqrt{\pi ^2 G_{\boldsymbol{q} \nu}^2 \left(\left((\hbar \beta \omega_{\boldsymbol{q}\nu})^2+8\right) \sinh \left(\frac{\hbar \beta \omega_{\boldsymbol{q}\nu}
   }{2}\right)-4 \hbar \beta \omega_{\boldsymbol{q}\nu}  \cosh \left(\frac{\hbar \beta \omega_{\boldsymbol{q}\nu} }{2}\right)\right)+4 \sinh
   \left(\frac{\hbar \beta \omega_{\boldsymbol{q}\nu} }{2}\right)}} \label{eq:gamma-approx}
\end{align}
\end{widetext}
We then directly compare~\cref{eq:tau-eph-2,eq:tau-eeph-2} and as such we simplify the expression under the square root of~\cref{eq:gamma-approx} by taking its power series around $G_{\boldsymbol{q} \nu}=0$ to first order to obtain:
\begin{align}
   \gamma( \hbar \beta \bar{\omega}_{\boldsymbol{q}\nu}) &\approx  \frac{\hbar \beta \omega_{\boldsymbol{q}\nu}}{4 G_{\boldsymbol{q} \nu} \sinh^2(\hbar \beta \omega_{\boldsymbol{q}\nu}/2)} \label{eq:gamma-approx-2}
\end{align}
Substituting~\cref{eq:gamma-approx-2} into~\cref{eq:tau-eeph-2}, we see that the two rates are identically equal.
As such, we conclude that any additional (positive) scattering terms arising from terms higher than linear order in $G_{\boldsymbol{q} \nu}$ can only decrease the phonon-mediated electron-electron lifetime, showing that $\tau_{\rm ee(ph)}$ is strictly smaller than $\tau_{\rm e-ph}$.

\Cref{eq:gamma-approx,eq:gamma-approx-2} are most valid for dimensionless electron-phonon couplings $G_{\boldsymbol{q} \nu} \ll 1$, which is true for non-superconducting materials systems.
Further, the phonon-mediated electron-electron interaction proceeds via the exchange of a `virtual' phonon, i.e. the phonon emitted (absorbed) by a pair of electrons is assumed to be instantaneously absorbed (emitted) by a different pair of electrons.
In practice, this means that other phonon scattering mechanisms, such as anharmonic phonon-phonon scattering, must be `quenched' out, that is, occur at a much slower rate.
This assumption is justified at low temperatures, but breaks down at temperatures approaching the material's Debye temperature.
At those elevated temperatures, one can still use~\cref{eq:gamma-approx} by augmenting $G_{\boldsymbol{q} \nu}$ with a competing coupling $G^{pp}_{\boldsymbol{q} \nu}$ capturing anharmonic phonon-phonon interactions.
Integrating~\cref{eq:tau-eeph-2} numerically using this competing coupling, we find that $\tau_{\rm ee(ph)}$ naturally increases, and can overtake $\tau_{\rm eph}$ at high temperatures.
In both ZrSiS and TaAs$_2$, (as well as WP$_2$ and WTe$_2$), there is no energy separation between the acoustic branches and the lowest-energy optical modes. 
These lower-energy optical modes involve large displacements of the heavy Zr/Ta atoms. 
The heavy transition metal atoms will result in a narrow bandwidth of the acoustic branches and thus a smaller phase space and long phonon-phonon lifetimes, as discussed in high thermal-conductivity system cubic boron arsenide~\cite{Lindsay2013first}.

This analytical limit allows us to revisit and further establish our design principles discussed in Sec.\cref{sec:overview_materials}.
To obtain a low enough $l_{\mathrm{mc}}$, we need a material with non-vanishing density of states at the Fermi level, and a large electron-phonon matrix element. 
While the former observation encourages us to seek for short $\tau_{\rm ee(ph)}$ in metals, the latter indicates that the electron potential is strongly sensitive to lattice perturbations, i.e. phonons. 
This is the case when the atomic orbital mixing has low symmetry (i.e. the band extrema are located off high symmetry points in the Brillouin zone), where the electron wavefunction, or equivalently the charge distribution in real space, is highly anisotropic.
To maintain a long enough $l_{\rm mr}$, large Fermi velocity facilitated by dispersive electronic bands can be combined with long electron-phonon lifetime maintained by reduced electronic phase space at low temperatures. 
Indeed, the electronic structures of both ZrSiS and TaAs$_2$ explored in this work, as well as WTe$_2$ and WP$_2$ reported earlier, display the above features.
There, the Fermi surfaces arise from a combination of $d$ orbitals from the transition metal and $p$ orbitals from the metalloids.
These material-specific insights are poised to inform exploration hydrodynamic electron flow in various condensed matter systems.

\section{Conclusions and Outlook}
This \emph{Article} identifies material signatures for hydrodynamic electron flow in semimetals from first principles, and predicts two candidates with prominent hydrodynamic effects across a wide temperature range that warrant experimental exploration.
We obtain the momentum-relaxing electron-phonon mean free paths $l_{\rm mr}$ and the momentum-conserving phonon mediated electron-electron mean free paths $l_{\rm mc}$ in ZrSiS and TaAs$_2$, and show that the indirect phonon-mediated electron-electron interaction could dominate over a wide temperature range, facilitating hydrodynamic behavior.
Through inspection of the momentum relaxing lifetimes on their Fermi surfaces, we find that at low temperatures, hole pockets with open Fermi surface shape feature much longer lifetimes than electron pockets.
These observations suggest that topological Dirac/Weyl bands are not indispensable in the search for promising candidates for hydrodynamic flow, but highly dispersive bands are beneficial in realizing long $l_{\rm mr}$.
We review these signatures in light of previously studied hydrodynamic candidates and discuss the analytical limits of the first- and second-order electron-phonon interactions to find that the mechanism is more broadly applicable.
At the Eliashberg level of theory, we then propose general principles for experimental discovery of hydrodynamic behavior in anisotropic metals, such as low symmetry crystals with $d/p$ atomic orbital mixing, suppressed phonon-phonon scattering, and reduced electronic phase space.
Future work distinguishing electron-electron Umklapp scattering events and computing the electron viscosity tensor from first principles is still needed to provide a full picture of electron hydrodynamics in quantum materials.  

\section{Acknowledgement}

The authors acknowledge fruitful discussions with Adam Jermyn, Uri Vool, Doug Bonn, Ken Burch, and Amir Yacoby.
This work (Y.W., G.V., P.N.) was primarily supported by the Quantum Science Center (QSC), a National Quantum Information Science Research Center of the U.S. Department of Energy (DOE). P.N. is a Moore Inventor Fellow and gratefully acknowledges support through Grant No. GBMF8048 from the Gordon and Betty Moore Foundation. This research used resources of the Oak Ridge Leadership Computing Facility, which is a DOE Office of Science User Facility supported under Contract DE-AC05-00OR22725 as well as the resources of the National Energy Research Scientific Computing Center, a DOE Office of Science User Facility supported by the Office of Science of the U.S. Department of Energy under Contract No. DE-AC02-05CH11231.

\bibliographystyle{apsrev4-2}
\bibliography{citations}
\end{document}